\documentclass[fleqn,usenatbib, useAMS]{mnras}
\usepackage{graphicx}
\usepackage{multicol}
\usepackage{enumerate}
\usepackage{amsmath}
\usepackage{amssymb}
\usepackage[flushleft]{threeparttable}

\usepackage{journal_names}

\title[Candidate H$\alpha$ outliers towards the Galactic bulge] {Candidate H$\alpha$ emission and absorption line sources in the Galactic Bulge Survey}
\author[Wevers et al.]{T. Wevers$^{1}$\thanks{Email: t.wevers@astro.ru.nl}, P. G. Jonker$^{2,1}$, G. Nelemans$^{1,3}$, M. A. P. Torres$^{2,1}$, P. J. Groot$^{1}$, \newauthor D. Steeghs$^{4}$, T. J. Maccarone$^{5}$, R. I. Hynes$^{6}$, C. Heinke$^{7}$, C. Britt$^{6,8}$\\\\
$^{1}$Department of Astrophysics/IMAPP, Radboud University, P.O. Box 9010, 6500 GL Nijmegen, The Netherlands\\
$^{2}$SRON, Netherlands Institute for Space Research, Sorbonnelaan 2, 3584 CA Utrecht, The Netherlands\\
$^{3}$Institute for Astronomy, KU Leuven, Celestijnenlaan 200D, 3001 Leuven, Belgium\\
$^{4}$Department of Physics, University of Warwick, Coventry CV4 7AL, UK\\
$^{5}$Department of Physics and Astronomy, Texas Tech University, Box 41051, Lubbock, TX 79409-1051, USA\\
$^{6}$Department of Physics and Astronomy, Louisiana State University, Baton Rouge, LA 70803-4001, USA\\
$^{7}$Department of Physics, University of Alberta, CCLS 4-183, Edmonton, AB T6G 2E1, Canada\\
$^{8}$Department of Physics and Astronomy, Michigan State University, 5678 Wilson Road, Lansing, MI 48824, USA
}

\begin{document}
\date{Accepted 2016 November 28. Received 2016 November 25; in original form 2016 October 14}
\pagerange{\pageref{firstpage}--\pageref{lastpage}} \pubyear{2016}
\maketitle
\label{firstpage}

\begin{abstract}
We present a catalogue of candidate H$\alpha$ emission and absorption line sources and blue objects in the Galactic Bulge Survey (GBS) region. We use a point source catalogue of the GBS fields (two strips of (l\,$\times$\,b)\,=\,(6$^{\circ}\times1^{\circ}$) centred at b\,=\,1.5$^{\circ}$ above and below the Galactic centre), covering the magnitude range 16 $\leq$ $r^{\prime} \leq$ 22.5. We utilize ($r^{\prime}$\,--\,$i^{\prime}$, $r^{\prime}$\,--\,H$\alpha$) colour-colour diagrams to select H$\alpha$ emission and absorption line candidates, and also identify blue objects (compared to field stars) using the $r^{\prime}$\,--\,$i^{\prime}$ colour index. 
We identify 1337 H$\alpha$ emission line candidates and 336 H$\alpha$ absorption line candidates. These catalogues likely contain a plethora of sources, ranging from active (binary) stars, early-type emission line objects, cataclysmic variables (CVs) and low-mass X-ray binaries (LMXBs) to background active galactic nuclei (AGN). The 389 blue objects we identify are likely systems containing a compact object, such as CVs, planetary nebulae and LMXBs. Hot subluminous dwarfs (sdO/B stars) are also expected to be found as blue outliers. Crossmatching our outliers with the GBS X-ray catalogue yields sixteen sources, including seven (magnetic) CVs and one qLMXB candidate among the emission line candidates, and one background AGN for the absorption line candidates. One of the blue outliers is a high state AM CVn system. Spectroscopic observations combined with the multi-wavelength coverage of this area, including X-ray, ultraviolet and (time-resolved) optical and infrared observations, can be used to further constrain the nature of individual sources.
\end{abstract}


\section{Introduction}
\label{sec:introduction}
The presence of an ionising radiation field can lead to hydrogen emission lines, while the presence of neutral hydrogen can result in absorption features in the optical spectrum of astronomical objects. From the properties of the H Balmer lines one can infer characteristics of the system under study. For example, the properties of single and/or double-peaked lines can allow us to infer geometrical properties \citep{Horne1986}, the presence or absence of an accretion disc \citep{Schwope2000,Ratti2012}, or the nature of the compact object and/or donor star \citep{Steeghs2002, Spaandonk2010, Casares2015, Casares2016}.

Historically, large scale photometric H$\alpha$ surveys with modest spatial resolution focussed on extended sources of emission to study star-forming regions, galaxy groups and supernova remnants (\citeauthor{Davies1976} \citeyear{Davies1976}, and references there-in). More recently, higher resolution surveys such as the INT Photometric H$\alpha$ survey (IPHAS; \citeauthor{Drew2005} \citeyear{Drew2005}, \citeauthor{Barentsen2014} \citeyear{Barentsen2014}) have focussed on the Galactic plane to uncover and study compact sources of emission, typically associated with various stages of stellar evolution. 
The most noteworthy H$\alpha$ survey covering the Galactic bulge is the photographic Anglo-Australian Observatory UK Schmidt Telescope Supercosmos H$\alpha$ Survey \citep{Parker2005}, going down to R\,$\sim$\,19.5 mag in the latitude range $\vert$b$\vert \leq10^{\circ}$. Currently ongoing is the VST Photometric H$\alpha$ Survey of the Southern Galactic Plane and Bulge (VPHAS+, \citeauthor{Drew2014} \citeyear{Drew2014}) which will cover the Galactic bulge and plane in 5 filters down to at least 20$^{th}$ magnitude. 

The analysis of colour-colour diagrams (CCDs) to search for H$\alpha$ emission line objects has been introduced by the IPHAS collaboration. \citet{Witham2008} present a method and first results of this effort. A variety of source classes, including CVs (see also \citeauthor{Witham2008} \citeyear{Witham2006}, \citeyear{Witham2007}), early-type emission line stars \citep{Corradi2008,Corradi2010, Drew2008}, active late-type stars, young stellar objects \citep{Vink2008} and planetary nebulae \citep{Viironen2009, Sabin2010} have been identified. For an example of the expected source classes and their location in the CCD we refer to fig. 1 in \citet{Corradi2008}. 

The presence of an ultraviolet or X-ray photon field can ionise hydrogen atoms in its direct environment, and hence lead to an H$\alpha$ emission line in the optical spectrum. Binary systems containing a compact object, such as a white dwarf (WD), neutron star (NS) or black hole (BH), are  examples of (transient) H$\alpha$ emitters, with the strength and width of the emission line depending on the primary mass, mass accretion rate and inclination angle with respect to the line of sight \citep{Casares2015}. Other excitation mechanisms (e.g. collisional excitation in the stellar corona) can also excite H atoms and induce spectral line emission.\\
In addition to H$\alpha$ emission line objects, some systems show H$\alpha$ in absorption. If the strength of this absorption line is stronger than that of a normal main sequence (MS) star, it will appear as an outlier below the locus of stars in a CCD. For example, single H-rich (DA) WDs are known to exhibit a very broad H$\alpha$ absorption line. C-rich and S-type asymptotic giant branch (AGB) stars have molecular ZrO absorption bands in their spectra that coincide with the location of the H$\alpha$ line (e.g. ZrO\,$\lambda$6456). These objects will hence also appear to have a deficit of flux in the H$\alpha$ filter relative to the $r^{\prime}$-band and appear as outliers to the locus of objects which do not exhibit these features in their spectrum. Late-type variable stars can cover a large range in colour space depending on the relative strength of molecular absorption bands (e.g. TiO, VO, ZrO) that are located in the $r^{\prime}$, $i^{\prime}$ and H$\alpha$ filters. 

In this work, we use the optical observations from \citet{Wevers2016}, taken as part of the Galactic Bulge Survey (GBS; \citeauthor{Jonker2011} \citeyear{Jonker2011}, \citeyear{Jonker2014}), to search for sources with excess H$\alpha$ emission and absorption signatures compared to normal MS stars. We also identify blue outliers with respect to field stars in the CCDs.
The structure of this article is as follows: Section \ref{sec:methods} outlines the method used to identify outliers, and in Section \ref{sec:results} we present the results. We discuss our findings in Section \ref{sec:discussion} and summarise in Section \ref{sec:summary}.

\section{Data}
\label{sec:data}
\subsection{Photometry}
As the starting point of our work, we use the optical point source catalogue covering the GBS fields \citep{Wevers2016}. This catalogue consists of optical photometry obtained using the Mosaic-2 camera on the Victor M. Blanco telescope, located at the Cerro Tololo Inter-American Observatory (CTIO), in three filters: $r^{\prime}$, $i^{\prime}$ and H$\alpha$. The areas covered are centred on b\,=\,1.5$^{\circ}$ above and below the Galactic centre, and consist of two strips spanning (l\,$\times$\,b)\,=\,(6$^{\circ}\times1^{\circ}$). In total, 64 fields, each consisting of 8 frames, were observed twice. One of the two exposures was offset by $\sim$1.2 arcmin in right ascension and declination to fill the gaps between the detectors. Therefore we have 1024 observed frames in total. The mean 5$\sigma$ limiting depth of the observations is $r^{\prime}$\,=\,22.5, $i^{\prime}$\,=\,21.1 mag. The point source catalogue includes objects that have been detected with a signal-to-noise ratio of more than 5 in all bands. For more details about the optical catalogue we refer the reader to \citet{Wevers2016}. 

\subsubsection{Global photometric calibration}
\label{sec:calibration}
The observations consist of two strips of overlapping fields above and below the Galactic Centre, respectively (see fig.1 in \citeauthor{Wevers2016} \citeyear{Wevers2016}). We use the overlap between these observations to apply a photometric calibration with the goal of getting all the photometry on the same absolute scale. There is no overlap between northern and southern fields, so we calibrate them independently. To this end, we use the method developed by \citet{Glazebrook1994} (see also \citeauthor{Barentsen2014} \citeyear{Barentsen2014} for an application of this method to the IPHAS dataset). 

In short, the goal is to minimize the magnitude offsets between stars that are present on overlapping fields using a set of anchor fields for which the photometry is thought to be well determined. The reference fields are chosen on 2 photometric nights, namely MJD 53912 and 59315 (see table 1 in \citeauthor{Wevers2016} \citeyear{Wevers2016} and Section \ref{sec:quality}). We denote the magnitude offset between stars on overlapping fields as $\Delta_{ij}$\,=\,$\langle m_i - m_j \rangle$. In order to keep the solution from drifting arbitrarily far from the values of the anchor fields, the difference in zeropoint values across the fields is also minimized. This problem can be solved as a linear least-squares problem because the magnitudes and the zeropoints are linearly related. Following \citet{Glazebrook1994} and setting the weights $w_{ij}$\,=\,1, we minimize the sum: 

\begin{equation}
\label{eq:leastsquares}
S = \sum\limits_{i=1}^N \sum\limits_{j=1}^N \theta_{ij} (\Delta_{ij} + a_i - a_j)^2
\end{equation}
where $\theta_{ij}$ is an overlap function that is 1 when there is overlap and 0 otherwise, a$_i$ are the zeropoints to solve for and a$_j$ the zeropoints of overlapping fields to a$_i$. N is the total number of fields included in the least-squares problem.

Minimising the sum in equation \ref{eq:leastsquares} by varying a$_i$ is equivalent to solving $\frac{\delta S}{\delta a_i}$\,=\,0, which yields the matrix equation 
\begin{equation}
\sum\limits_{j=1}^N A_{ij} a_j = b_i
\end{equation}
where
\begin{equation}
 A_{ij} = \theta_{ij}\ - \delta_{ij} \sum\limits_{k=1}^N \theta_{jk}
 \end{equation}
 and 
 \begin{equation}
 b_i = \sum\limits_{j=1}^N \theta_{ij} \Delta_{ij}
\end{equation}

We now solve the least-squares problem by keeping the solutions of the anchor fields fixed, while the zeropoints of the other fields are allowed to vary to optimize the solution as a global photometric calibration. 

\subsubsection{Colour-colour diagrams}
\label{sec:ccd}
Before we move to the details of the selection criteria for outliers, we first introduce the necessary tools we will use to find them: colour-colour diagrams and synthetic photometry. We merge the nominal and offset observations in each filter (taken within minutes of each other), and create ($r^{\prime}$\,--\,$i^{\prime}$,$r^{\prime}$\,--\,H$\alpha$) colour-colour diagrams. This gives us a total 512 frames which form the basis of our work. We exclude sources with saturated photometry in our analysis. We use the same basic techniques as presented by the IPHAS collaboration \citep{Drew2005, Witham2006}.

We use a set of synthetic spectra \citep{Pickles1998} to create synthetic photometry for spectral types ranging from O5V to M5V using the CTIO filters\footnote{http://svo2.cab.inta-csic.es/svo/theory/fps3/index.php?mode\\=browse$\&$gname=CTIO}. We redden these spectra with increasing $E(B-V)$ to estimate the colours of stars at a range of reddening values (hence distances).
We consider solar-metallicity MS and giant stars. The binning of the spectra is sufficiently small (5\,\AA) that we can use them to compute synthetic photometry for our $r^{\prime}$ and $i^{\prime}$ filters as well as for the narrow-band H$\alpha$ filter. We recompute the grids of these filter profiles to match the binning of the spectra, meaning that for each spectral bin we compute the filter transmission value at the midpoint of the bin.
We define the synthetic colours in the Vega system as
\begin{center}
\large
\begin{equation}
m_1 - m_2 = -2.5\ \text{log } \frac{\int T_{1,\lambda} F_{\lambda} d\lambda}{\int T_{1,\lambda} F_{\lambda, \text{V}} d\lambda} + 2.5\ \text{log} \frac{\int T_{2,\lambda} F_{\lambda} d\lambda}{\int T_{2,\lambda} F_{\lambda, \text{V}} d\lambda}
\end{equation}
\normalsize
\end{center}
where the filter transmission profiles are labeled $T_x$, $F_{\lambda}$ is the synthetic spectrum per spectral type and $F_{\lambda, \text{V}}$ is the spectrum of Vega \citep{Bohlin2007}.
It is possible to compute an upper limit to the $r^{\prime}$--H$\alpha$ colour of any physical object based on synthetic photometry of a pure H$\alpha$ emission line spectrum \citep{Drew2005}. The upper limit is $r^{\prime}$--H$\alpha$\,=3.3 for the used filter combinations, hence we discard all objects that have an observed $r^{\prime}$\,--\,H$\alpha$ colour above this value. Such values are indicative of a bad crossmatch between the three optical bands, or detector artifacts.

\begin{figure*} 
  \includegraphics[height=13cm, keepaspectratio]{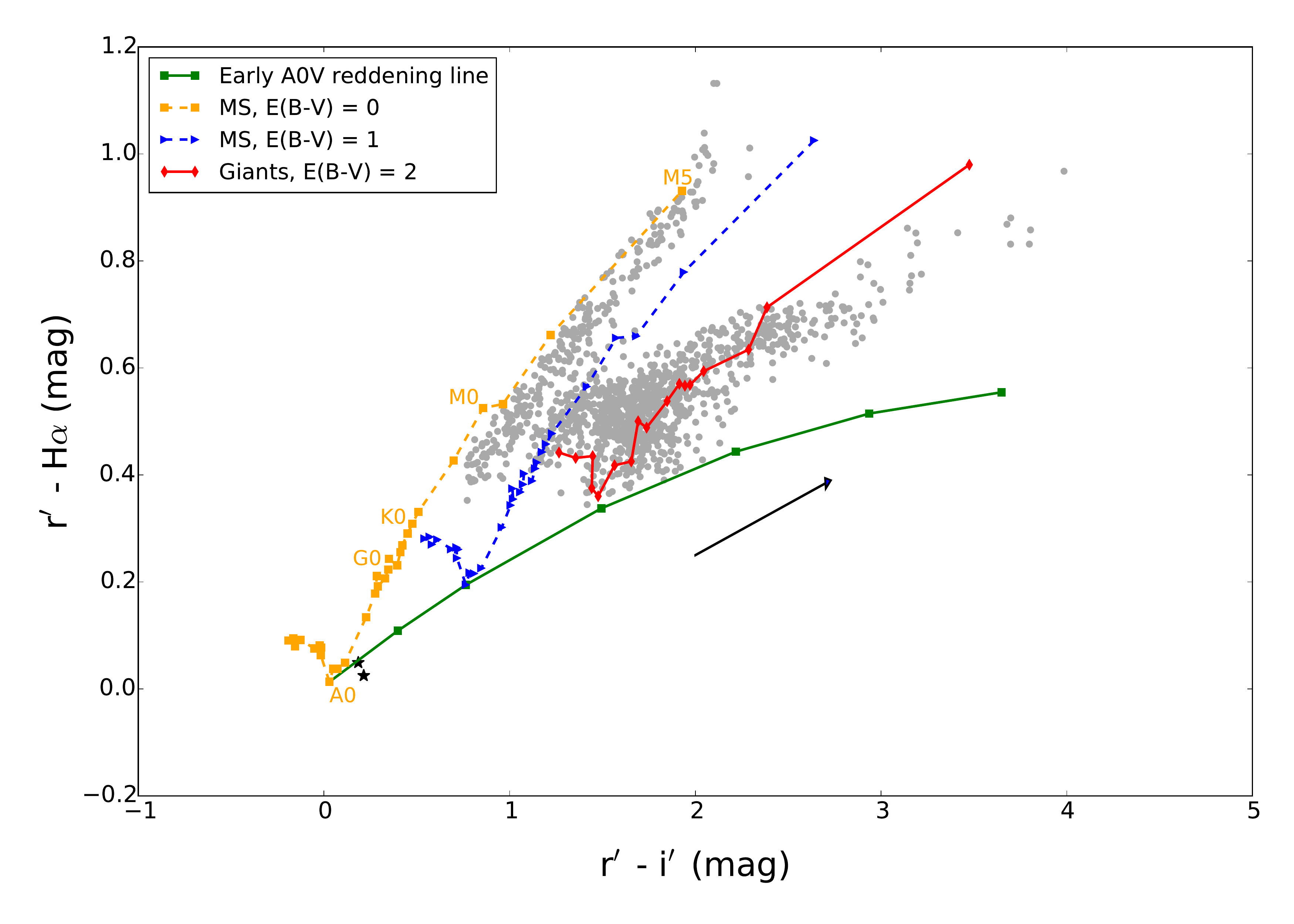}
  \caption{The black points show the observed colours of sources on field S01 (detector 2). The dashed lines are the spectral sequences of unreddened MS stars from O5V to M5V, with $E(B-V)$\,=\,0 and 1 for orange squares and blue triangles, respectively. The red diamonds show colours of giants with spectral types ranging from O8III to M5III, reddened to $E(B-V)$\,=\,2. The green squares show the early A-type MS line, which indicates the lower limit for A0V MS stars in the CCD. The arrow indicates the effect of reddening $\Delta$$E(B-V)$\,=\,1. The two outliers marked by black stars around (0.1,0) are very blue compared to the other field stars, making them potentially interesting sources.}
  \label{fig:ccdexample}
\end{figure*}

In Figure \ref{fig:ccdexample} we show an example of a CCD of field S01 (detector 2), centred at Galactic coordinates ($l, b$)\,=\,(--2.81, --1.75). Observed colours are plotted in black, and overlaid are synthetic tracks for MS stars (orange squares and blue triangles with $E(B-V)$\,=\,0 and 1, respectively) and giants (red diamonds, $E(B-V)$\,=\,2). The two main populations of stars that can be identified are the unreddened MS stars (located slightly to the right and below the orange synthetic track) and the locus of reddened stars. We note that in the CCD there is an offset between the synthetic track and the observed unreddened MS. This offset shows that the unreddened MS in this case is nevertheless slightly reddened to about $E(B-V)$\,$\leq$\,0.5. 

Comparing the observed CCD and the unreddened synthetic track reveals that our catalogue apparently contains no unreddened stars of early spectral types. The observed unreddened MS population typically contains only stars of spectral type K0V or later. Taking the absolute magnitude from \citet{Schmidtkaler1982} and colour from \citet{Pecaut2013}, and the colour transformation given by \citet{Jester2005}, we find that a K0V star observed at $r^{\prime}$\,=\,17 (the typical saturation limit of the optical catalogue) is located at a distance of $\sim 1$ kpc. Unreddened stars with a spectral type earlier than K0V (hence intrinsically brighter) and located within 1 kpc are saturated. The photometric observations of saturated sources are unreliable and we discard them in our analysis. 

\begin{figure} 
  \includegraphics[height=5.5cm, keepaspectratio]{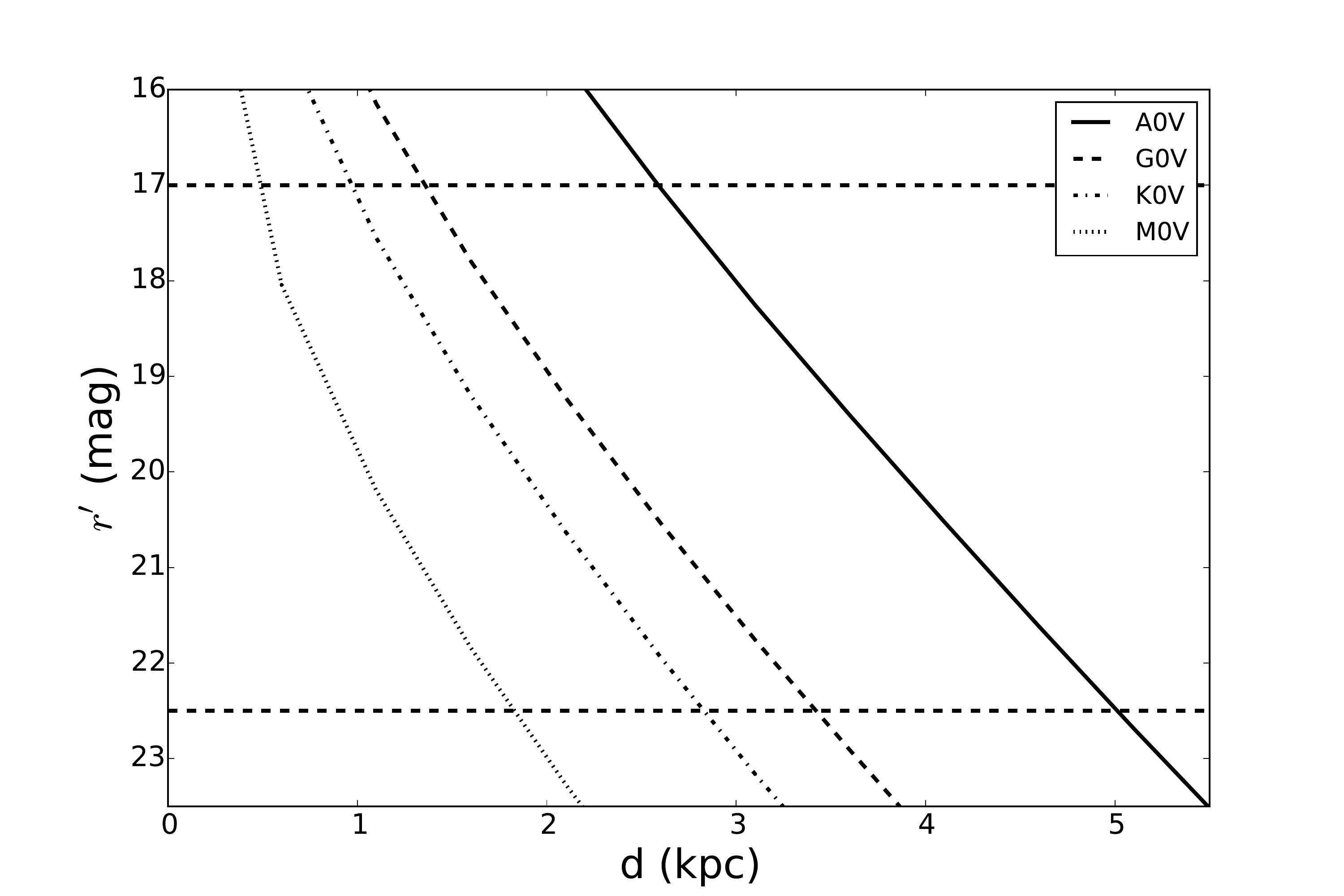}
  \caption{Estimated distance ranges for MS stars of different spectral types present in the optical GBS catalogue, assuming a saturation limit of $r^{\prime}$\,=\,17 (upper horizontal dashed line) and a limiting magnitude of $r^{\prime}$\,=\,22.5 (lower horizontal dashed line). See Section \ref{sec:ccd} for details. }
  \label{fig:limits}
\end{figure}

We use the absolute magnitudes from \citet{Schmidtkaler1982} together with the 3D reddening map from \citet{Schultheis2014} at ($l$, $b$) = (-2.8, -1.8) to estimate the distance ranges for stars of different spectral types in our catalogue. The 3D reddening map is converted to the $r^{\prime}$-band following \citet{Schlegel1998}. In Table \ref{tab:limits} we show the results for different spectral types. We use the distance modulus to estimate the observable distance range as: 
\begin{equation}
\text{log}\ d\ \text{(pc)}= 0.2 \times (r^{\prime} - M_{r^{\prime}} - A_{r^{\prime}}\text{(}d\text{)} + 5)
\end{equation}
where M$_{r^{\prime}}$ is the absolute magnitude, $r^{\prime}$ the apparent (observed) magnitude and A$_{r^{\prime}}$ the extinction in the $r^{\prime}$-band obtained from the \citet{Schultheis2014} reddening map. The results of this calculation are visualised in Figure \ref{fig:limits}. We assume that the saturation limit of our catalogue is $r^{\prime}$\,=\,17, and the limiting magnitude is $r^{\prime}$\,=\,22.5 (marked by dashed horizontal lines), giving rise to the ranges shown in Table \ref{tab:limits}. Using our synthetic photometry, we can also infer the range of $r^{\prime}$\,--\,$i^{\prime}$ colours which different spectral types occupy in the CCD, assuming that E($r^{\prime}$\,--\,$i^{\prime}$)\,=\,0.26\,$\times$\,A$_{r^{\prime}}$ \citep{Schlegel1998}. 

\begin{table}
 \centering
  \caption{Estimate of the distance range and observed colours of stars with different spectral types in field S01 (detector 2).}
  \begin{tabular}{cccccc}
  \hline
Sp. type & M$_{r^{\prime}}$ & ($r^{\prime}$\,--\,$i^{\prime}$)$_{\text{syn}}$ &  d (kpc) & A$_{r^{\prime}}$ & ($r^{\prime}$\,--\,$i^{\prime}$)$_{\text{obs}}$   \\\hline
A0V & 0.77 & 0 & 2.6 -- 5 & 4.4 -- 8.4 & 1.1 -- 2.2  \\ 
G0V & 4.26 & 0.39 & 1.3 -- 3.4 & 2.2 -- 5.7 &1.0 -- 1.9 \\
K0V & 5.67 & 0.48 & 0.9 -- 2.8 & 1.5 -- 4.7 & 0.9 -- 1.7 \\
M0V & 11.72 & 0.96 & 0.5 -- 1.8 & 0.8 -- 3.0 &1.2 -- 1.8 \\
  \hline
  \end{tabular}
  \label{tab:limits}
\end{table}

We conclude that the locus of reddened stars at $r^{\prime}$\,--\,$i^{\prime}$\,$\sim$\,1.4 consists of reddened MS stars with spectral type earlier than M0V. The objects located beyond $r^{\prime}$\,--\,$i^{\prime}$\,$\sim$\,2 are giants, as MS stars are too faint to be observable at those reddening values.

We briefly note that there are two outliers around ($r^{\prime}$\,--\,$i^{\prime}$, $r^{\prime}$\,--\,H$\alpha$)\,=\,(0.1,0). As we explained above, these can not be ordinary early-type MS stars because they would appear saturated in our observations. 

\subsection{Spectroscopy}
A 500\,s spectrum of the optical counterpart to the X-ray source CX2 \citep{Jonker2011} was taken on 2010 July 8 with the ESO Faint Object Spectrograph and Camera (EFOSC2, \citeauthor{Buzzoni1984} \citeyear{Buzzoni1984}) at the ESO New Technology Telescope (NTT). We used grism $\#$13 combined with a 1 arcsec~slit, resulting in a spectral resolution of R\,$\sim$\,300 and a wavelength coverage ranging from 3700\,--\,9300\,\AA. We debiased and corrected for the CCD flatfield response, and wavelength calibration was performed using a HeAr arc lamp. We normalised the spectrum by fitting cubic splines to the continuum in \textsc{molly}. 

\section{Outlier identification}	
\label{sec:methods}
We devised a selection method that results in the automatic identification of H$\alpha$ emission and absorption line candidates and blue sources (all referred to as outliers). Our dataset consists of 1024 observations, observed on 8 nights, spanning a large range of stellar densities, dust extinctions and photometric uncertainties. Hence it is unavoidable that any selection method will fail in some cases. We take a conservative approach and prioritize minimising false positives over completeness. This implies that our catalogue will not be complete. Below we introduce the automatic identification algorithm together with a list of criteria that must be fulfilled for the results to be deemed trustworthy. CCDs that fail to meet these criteria have been rejected from automatic processing and are instead inspected manually.

We set out to identify the outliers from the main features that are present in the CCD: the unreddened MS and reddened sequence. 
We make a distinction between these two features and fit them independently. We identify H$\alpha$ emission line sources from both the unreddened and reddened loci of objects. Additionally we identify absorption line sources from the reddened locus, which should be located below the main population in the CCD. Unreddened stars with strong absorption line features may overlap with more reddened objects (and conversely H$\alpha$ emitters from the reddened locus may overlap with the unreddened MS) and cannot be distinguished based on a CCD alone. Including a colour-magnitude diagram in the analysis may help to break this degeneracy, but this is beyond the scope of this work. 

We perform the analysis (described in detail below) for two magnitude bins, one including sources with $r^{\prime}$\,$\leq$\,19.5 and one containing sources with $r^{\prime}$\,$\geq$\,19.5. The motivation to use two magnitude bins is the fact that at fainter magnitudes, the photometric uncertainties increase and hence the scatter of stars in the CCD also increases. If we combine sources with small and large photometric errors in the same CCD, our selection criteria will be dominated by the intrinsic scatter of the faint sources. This may preclude us from identifying sources with small photometric uncertainties as significant outliers. We use the value $r^{\prime}$\,=\,19.5 because the peak of the distribution of magnitudes in the $r^{\prime}$-band typically occurs around this magnitude. It is approximately in the middle between the saturation limit and the 5$\sigma$ detection limit, and roughly the completeness limit of the optical catalogue \citep{Wevers2016}. 

\subsection{Outliers from the unreddened MS}
We define unreddened objects as sources that have $E(B-V)$ $\leq 1$ (i.e. all objects that lie above the synthetic track with $E(B-V)$\,=\,1, see Fig. \ref{fig:ccdexample}). As noted already in \citet{Witham2006}, fitting a straight line to this selection of objects may not converge onto the observed unreddened MS. The solution proposed by these authors is to iteratively force the fit upwards, and we do the same here. After an initial fit to all points with $E(B-V)$\,$\leq$\,1, we select the objects above the fitted line and iterate, forcing the fit up towards the unreddened MS. In practice, the shape of the CCD is determined by the detection limit of our observations, the stellar density and reddening along the line of sight. CCDs along different lines of sight have a different shape depending on these parameters, hence they require a different number of iterations for the fit to converge onto the unreddened MS. To establish whether or not our fit represents the unreddened MS, we calculate the slope of our fit in each iteration. If the slope of the fit is more shallow than the slope of the synthetic track between spectral types K5V and M5V, we deem our fit unsatisfactory and apply an additional iteration (i.e. we force the fit upward). We iterate for a maximum of 5 times; CCDs for which the fit has not converged at that point will be inspected manually for outliers. We also place a constraint on the $r^{\prime} - i^{\prime}$ colour of an unreddened MS star, to distinguish those sources from reddened stars with H$\alpha$ in emission. These sources may occupy the same parameter space in the CCD. However, $r^{\prime}$\,--\,$i^{\prime}$ increases for later spectral types. We conservatively estimate from the extent of the unreddened MS in our CCDs that the highest $r^{\prime}$\,--\,$i^{\prime}$ colour an unreddened late-type MS star can have is $r^{\prime}$\,--\,$i^{\prime}$\,=\,3.5.

Once we have identified the unreddened MS, we determine the iteratively 4$\sigma$-clipped scatter around the fit. We define outliers as sources with an excess H$\alpha$ emission contribution, quantified as follows: 
\begin{equation}
\label{eq:outlier}
(r^{\prime}-\text{H}\alpha)_{obs} - (r^{\prime}-\text{H}\alpha)_{fit} \geq C \times \sqrt{\sigma_s^2 + \sigma_{phot}^2}
\end{equation}
Here $\sigma_{s}$ represents the scatter of datapoints around the fit, and $\sigma_{phot}$ is the photometric measurement error for the $r^{\prime}$\,--\,H$\alpha$ colour index. C is a constant which we set to 4. An example is shown in Figure \ref{fig:CCDfit}. The resulting fit is overplotted as the upper solid line, while the dashed lines indicate the 4$\sigma$ scatter. Note that this is only a mean representation of the scatter, as it differs for each individual measurement. As an illustration we include the photometric uncertainties in $r^{\prime}$\,--\,H$\alpha$. 

\subsection{Outliers from the locus of reddened stars}
For the next step, we remove all sources belonging to the unreddened MS (defined as all datapoints that are within 4$\sigma$ of the final fit) and continue our analysis with the remaining objects. If the slope of our fit after 5 iterations is still more shallow than that of the synthetic photometry, we cannot identify the unreddened MS. In that case we remove all points above the synthetic track with $E(B-V)$\,=\,1, with the exception of sources that have $r^{\prime}$\,--\,$i^{\prime}$\,$\geq$\,3.5 (as these cannot be part of the unreddened MS).
We are now left with a sample of reddened stars, and continue to identify outliers by fitting a straight line to the remaining objects. 
As for the unreddened case, we determine the iteratively 4$\sigma$-clipped scatter to obtain the best fit and the scatter around it. For sources that are located below the locus of reddened objects (the absorption line sources) the left hand side of Equation \ref{eq:outlier} changes to an absolute value. Moreover, sources that have H$\alpha$ in absorption will have a comparatively lower signal-to-noise ratio in the H$\alpha$ filter (see Section \ref{sec:results}). It is expected that the scatter of these sources in the CCD will be larger than the typical scatter of the locus of stars. We therefore use a more conservative value of C\,=\,5 for the identification of absorption line candidates.

Visual inspection of the H$\alpha$ absorption line candidates shows that we identify many sources that fall partially off the detector when the H$\alpha$ filter is mounted but not when the $r^{\prime}$ filter is present. This slight shift in the focal plane is likely introduced by the different optical path the incoming light follows when the H$\alpha$ filter is mounted. To remove these spurious sources, we require that the source position is more than 20 pixels from the detector edges. 
\begin{figure*}
\minipage{0.5\textwidth}
  \includegraphics[width=\linewidth]{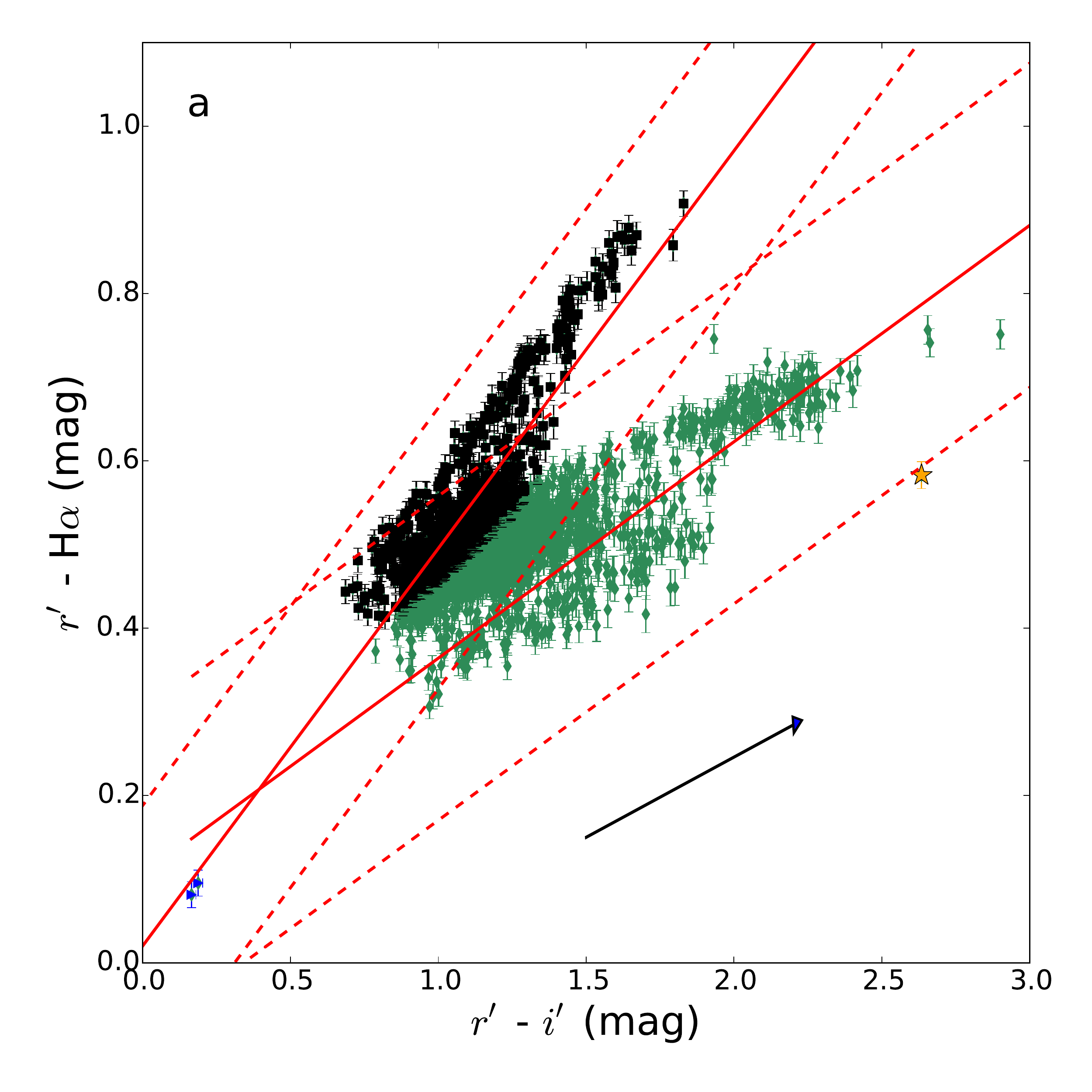}
\endminipage\hfill
\minipage{0.5\textwidth}
  \includegraphics[width=\linewidth]{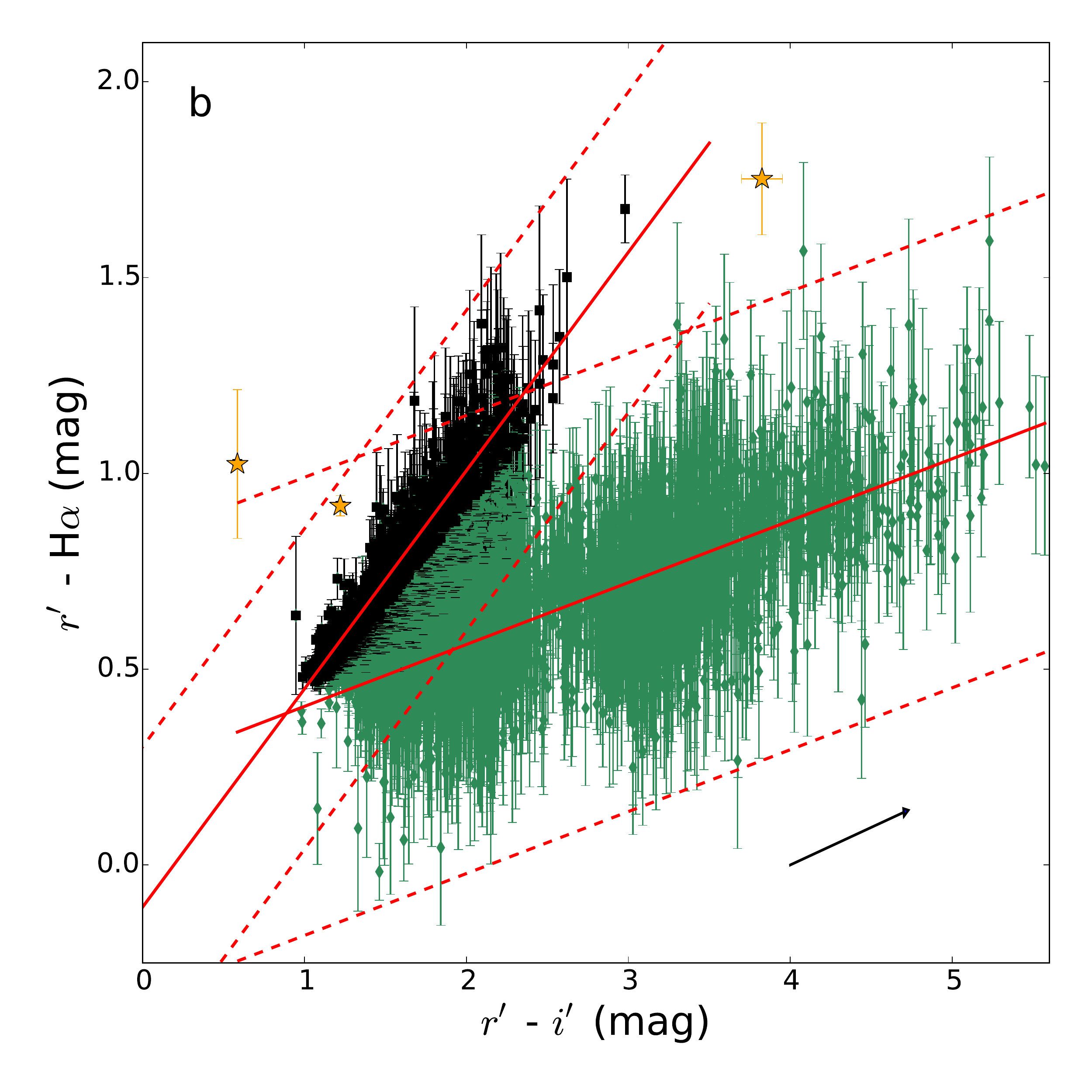}
\endminipage
\caption{Panel a: CCD of all stars brighter than $r^{\prime}$\,$\leq$\,19.5 on field S20, detector 6. The red solid lines indicate the fits to the loci of stars; dashed lines indicate the 4$\sigma$ scatter around these fits. Black squares are identified as part of the unreddened MS, green diamonds belong to the reddened locus of stars. Blue outliers are are plotted in blue, candidate H$\alpha$ outliers as gold stars. The reddening vector for $\Delta$$E(B-V)$\,=\,1 is shown as an arrow. Panel b: same, but for all stars fainter than $r^{\prime}$\,$\geq$\,19.5. }
\label{fig:CCDfit}
\end{figure*}
Figure \ref{fig:CCDfit} shows an example of two CCDs for the bright (panel a) and faint (panel b) magnitude bins. In particular panel b illustrates the diversity of photometric measurement errors (even for stars in the same magnitude bin), which can greatly influence outlier detection. The emission line candidate from the unreddened MS is inside the mean 4$\sigma$ boundaries, but still a significant outlier because of the very small photometric uncertainty. The emission line candidate at $r^{\prime}$\,--\,$i^{\prime}$\,=\,3.8 is also consistent with the expected locus of unreddened stars. However, the high $r^{\prime}$\,--\,$i^{\prime}$ colour implies it cannot be an unreddened object, so we detect it as an outlier to the locus of reddened sources.

\subsection{Blue outliers}
\label{sec:blue}
In Figure \ref{fig:CCDfit}, the fit to the unreddened MS (the upper solid line) is satisfactory in panel a, with a slope consistent with that of the unreddened MS synthetic photometry. There are two objects with very blue colours compared to the fields stars in the diagram (marked by blue squares). However, we do not identify them as outliers because they fall within the limits of the 4$\sigma$ regions around our fitted lines. Because these are potentially interesting sources, we use a different selection algorithm based on their $r^{\prime}$\,--\,$i^{\prime}$ colour. 

\begin{figure*} 
\minipage{0.5\textwidth}
  \includegraphics[width=\linewidth]{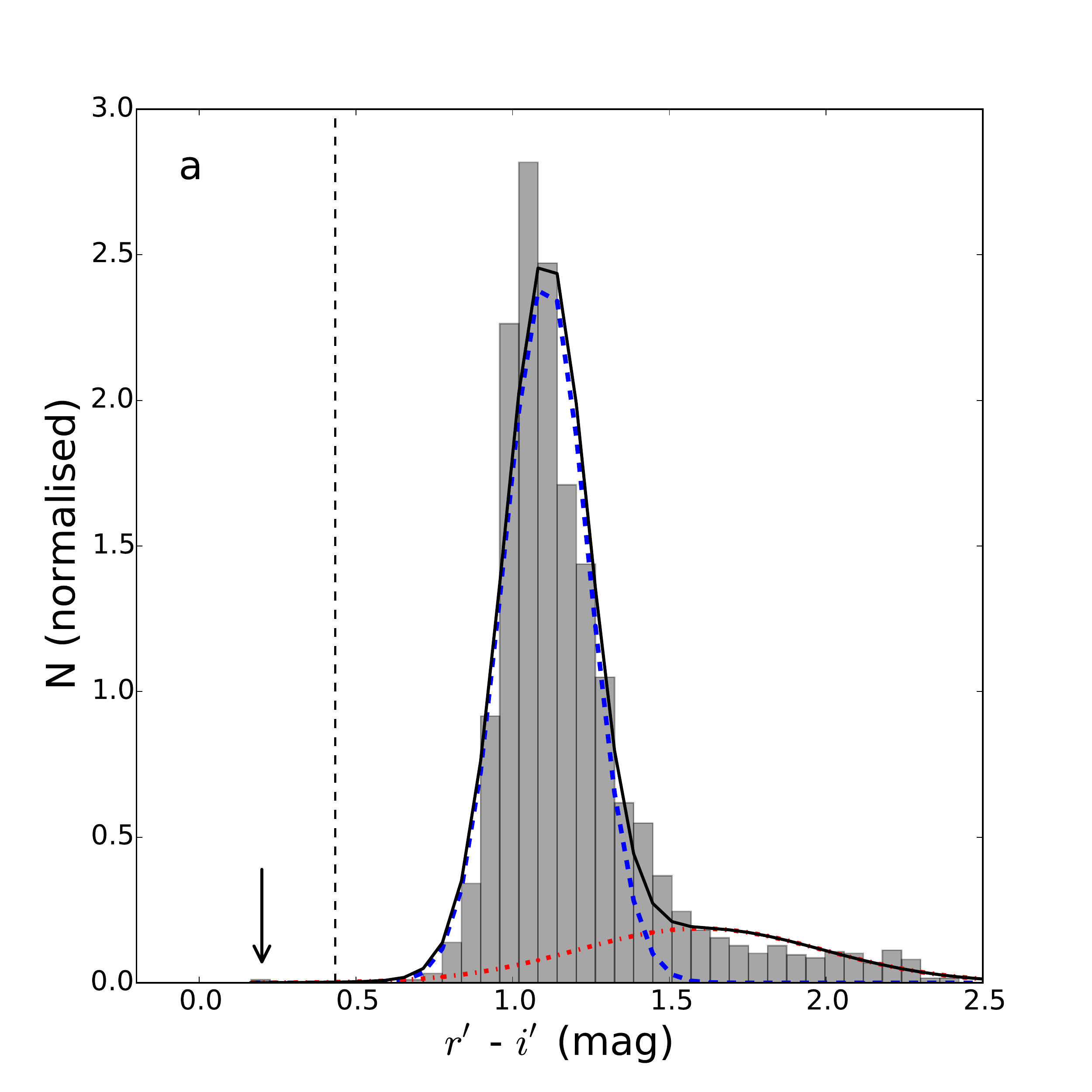}
\endminipage\hfill
\minipage{0.5\textwidth}
  \includegraphics[width=\linewidth]{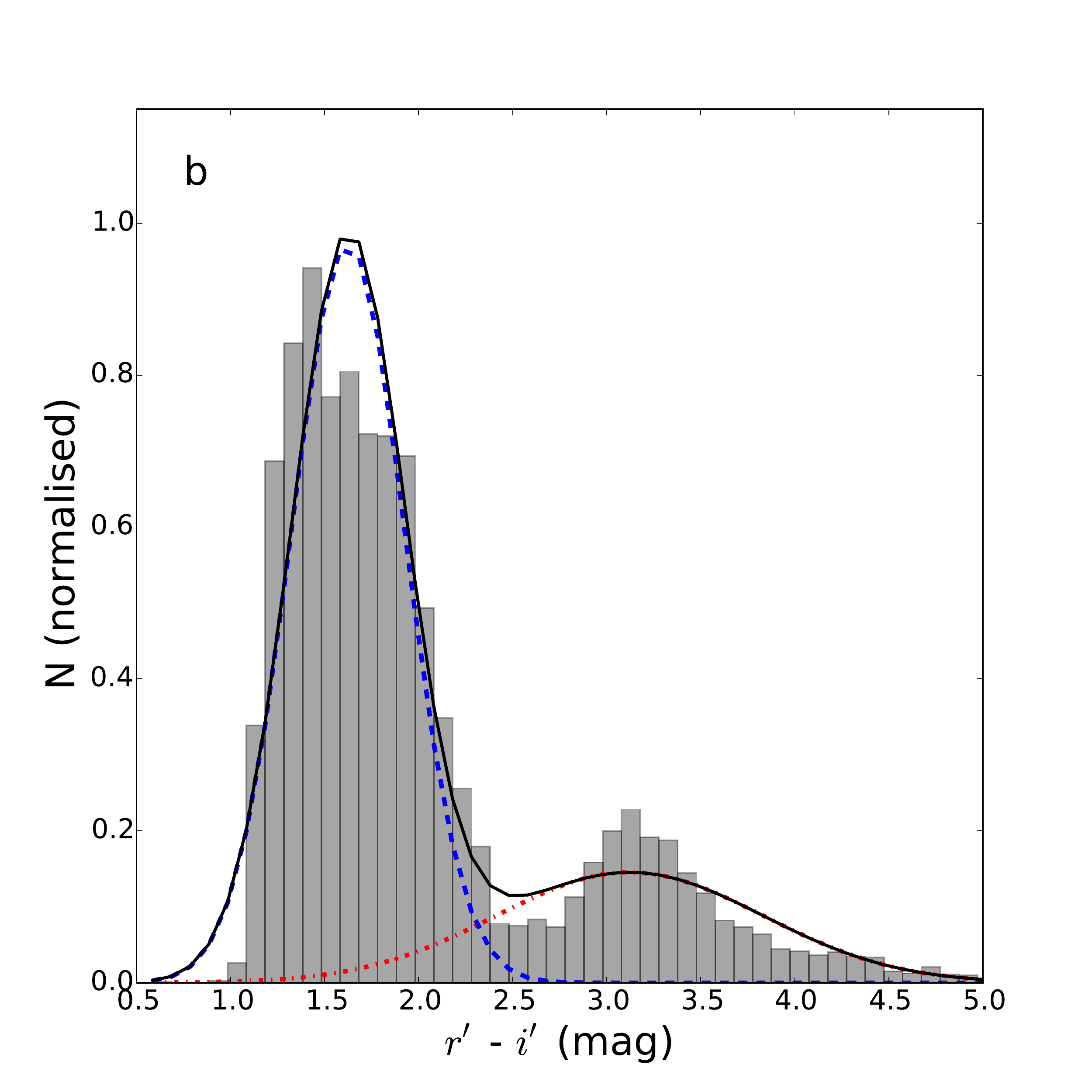}
\endminipage
\caption{Panel a: normalised $r^{\prime}$\,--\,$i^{\prime}$ colour distribution of all stars brighter than $r^{\prime}$\,$\leq 19.5$ on field S20, detector 6. The blue (dashed) and red (dash-dotted) lines show the Gaussian distributions; the black solid line shows the sum of both components. The vertical dashed line shows the 5$\sigma$ limit below which sources are flagged as outliers. There are two outliers around $r^{\prime}$\,--\,$i^{\prime}$\,=\,0.2, marked with an arrow. Panel b: same as panel a, but for all stars fainter than $r^{\prime}$\,$\geq 19.5$. No blue outliers are identified.}
\label{fig:Gaussfit}
\end{figure*}

We fit the distribution in $r^{\prime}$\,--\,$i^{\prime}$ with a Gaussian mixture model using the \textsc{python} Machine Learning package \textit{sci-kit learn} \citep{Pedregosa2011}. Based on the shape of the histogram, which is typically single- or double-peaked, we consider two models, one consisting of one Gaussian distribution and one consisting of the sum of two Gaussians. We note that there is no reason to believe that the blue edge of this distribution should follow a one-sided Gaussian distribution function. The shape of this colour distribution is determined by many factors, including the stellar density and magnitude distribution of stars along the line of sight, our survey detection limits and the effects of reddening. Typically the blue edge of the distribution is sharper than Gaussian (Carmona-Ruiz et al. in prep.), hence our approximation by a Gaussian distribution is a conservative approach. We show a result of fitting two Gaussian distributions to the number of stars as a function of their $r^{\prime}$\,--\,$i^{\prime}$ colours in Figure \ref{fig:Gaussfit} for the same field as in Figure \ref{fig:CCDfit}.

We determine the width, height and peak position using an iterative maximum likelihood estimate approach. Given the best fit parameters, we flag all sources more than 5$\sigma$ away from the peak of the distribution as outliers (marked in the figure by the dashed vertical line). In the case that the sum of two Gaussians works best, we select the blue (lowest $r^{\prime}$\,--\,$i^{\prime}$ peak position) Gaussian component to determine which sources are blue outliers. 

\subsection{Quality control}
\label{sec:quality}
The methods that we employ to find outliers do not rely on any underlying physical models that accurately predict the shapes or positions of the populations of sources we are trying to describe. We have optimised our methods such that they work for the majority of the frames, but visual inspection shows that in some cases our selection procedures do not yield satisfactory results. We therefore visually inspect every CCD to reject all anomalous frames. 

First of all, there are 75 frames for which the slope of our fit after five iterations is still more shallow than the slope of the synthetic track. As was mentioned earlier, this can arise due to a combination of the stellar density and reddening along the line of sight together with our selection criteria for (un)reddened objects. In 9 cases the iterative procedure reduced the number of stars available for the fit to only a handful. Visual inspection shows that no outliers were missed in these frames. In the remaining 66 frames the upper part of the population in the CCD is well identified, even though the slope is more shallow than our threshold value. The outliers we find in these frames are robustly identified and included in the final sample.

Secondly, the presence of H\,\textsc{ii} regions can affect the CCD by increasing the number of apparent H$\alpha$ emission line sources due to increased non-homogeneous extended emission. This is the case in fields S04 and N15. In \textsc{simbad} we find H\,\textsc{ii} regions LBN1120 and LBN9 for S04 and N15, respectively. 
The CCDs of field S07 contain a very large number of H$\alpha$ emission and absorption line outliers. We find an open star cluster in this field, containing many bright (V\,$\sim$\,6) stars that are saturated in our observations. We attribute the large number of outliers to the presence of these bright stars, which cause blooming of charge in the detectors that lead to false source detections and colours, and/or erroneous matches (see \citeauthor{Wevers2016} \citeyear{Wevers2016} for a discussion about the effect of saturated sources on e.g. source detection).

\begin{table*}
 \centering
   \caption{Example of the tables containing the information of the sources identified as outliers, including the position, magnitudes and colours. The photometric measurements are quoted in Vega magnitudes. }
  \begin{tabular}{cccccccccc}
  \hline
 RA ($^{\circ})$ & Dec $(^{\circ})$  &  $r^{\prime}$ & $\sigma_{r^{\prime}}$ &  $i^{\prime} $ & $\sigma_{i^{\prime}}$ & $H{\alpha}$ & $\sigma_{H{\alpha}}$ &$r^{\prime}$\,--\,$i^{\prime}$&$r^{\prime}$\,--\,H$\alpha$ \\
  \hline\hline
266.344818&  --32.464782 & 19.15&	0.02& 16.81 & 0.01 & 18.08 & 0.02 & 2.34 & 1.07 \\
266.251831&	--32.328826 &	17.32&	0.02& 15.50 & 0.01 & 16.19 & 0.02 & 1.82 & 1.13\\
266.208557&	--32.298542 &	19.01&	0.02& 17.00 & 0.01 & 18.05 & 0.02 & 2.01 & 0.96\\
266.442108&	--32.151733 &	19.31&	0.02& 17.37 & 0.01 & 18.37 & 0.02 & 1.94 & 0.94\\
266.280395&	--32.102737 &	17.96&	0.02& 16.71 & 0.01 & 16.80 & 0.02 & 1.25 & 1.16\\
  \hline
  \end{tabular}
  \label{tab:outliers}
\end{table*}

\section{Results}
\label{sec:results}
\subsection{Outlier populations}
\begin{figure*} 
\minipage{0.5\textwidth}
  \includegraphics[height=6cm, keepaspectratio]{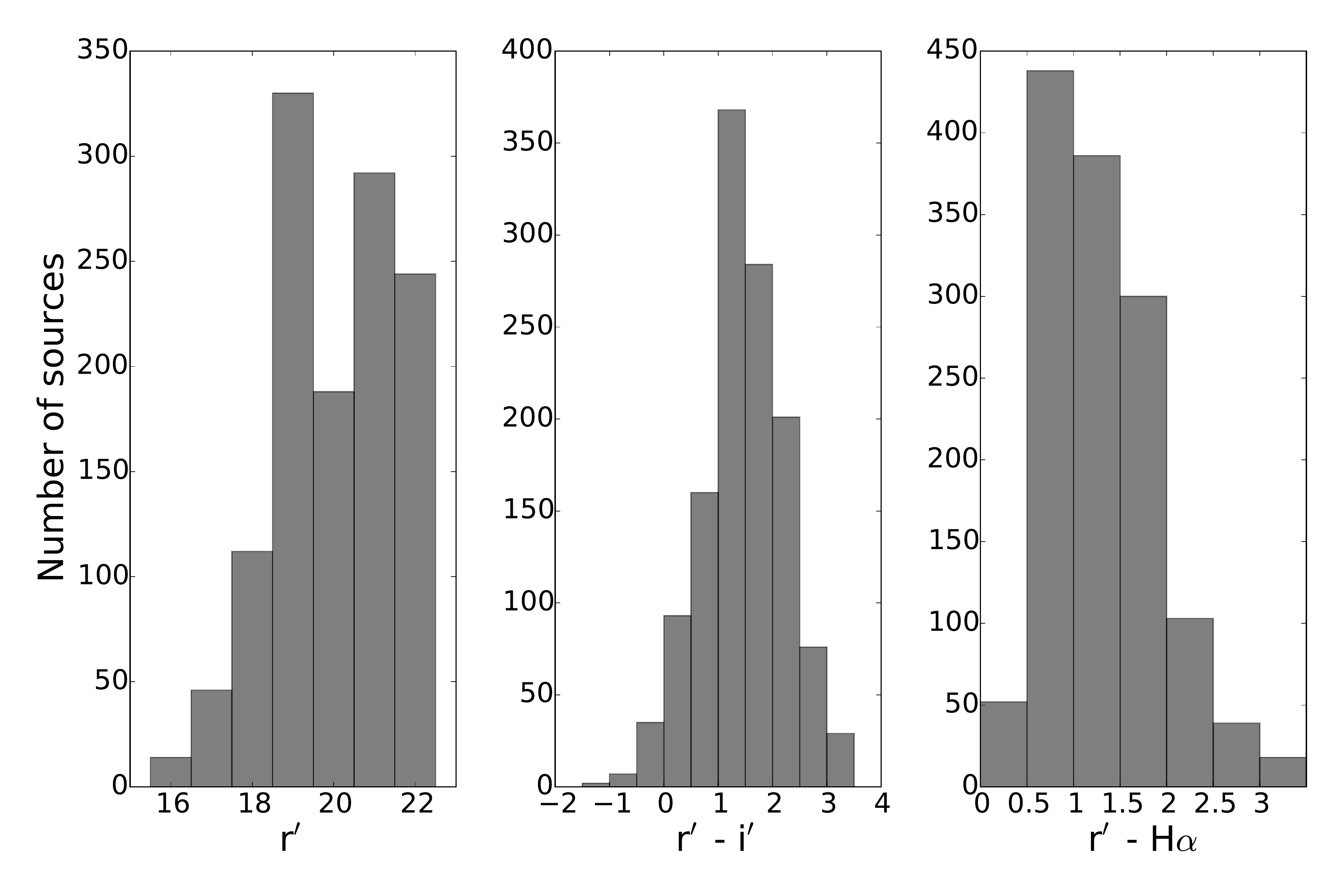}
\endminipage\hfill
\minipage{0.5\textwidth}
  \includegraphics[height=6cm, keepaspectratio]{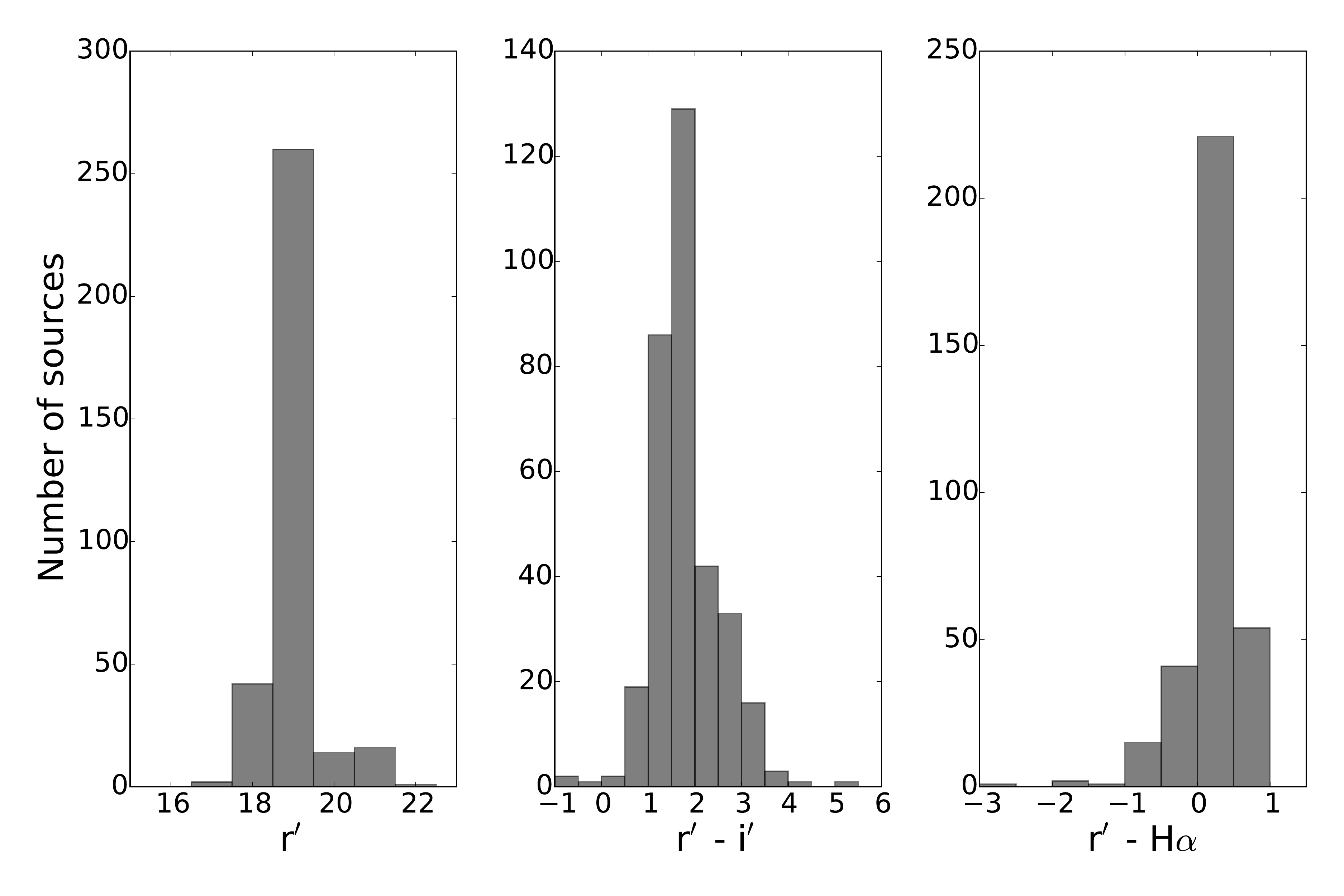}
\endminipage\hfill
\caption{Magnitude and colour distribution of the sample of candidate H$\alpha$ emission (left three panels) and absorption (right three panels) line outliers. An explanation for the decreased number of outliers in the magnitude bin 19.5\,$\leq r^{\prime}$\,$\leq$\,20 is given in the text.}
\label{fig:halphadist}
\end{figure*}

\begin{figure}
  \includegraphics[height=6cm, keepaspectratio]{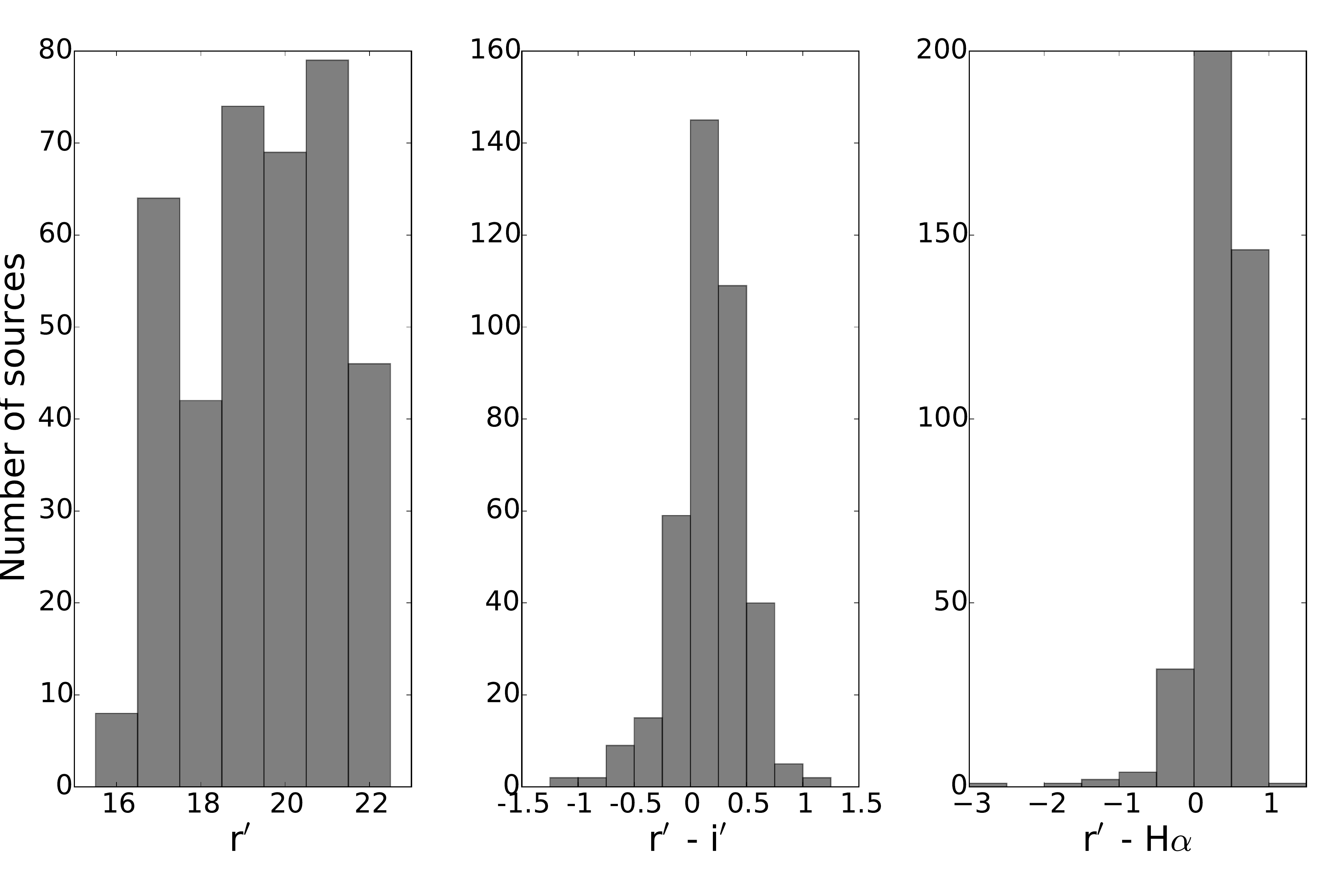}
  \caption{Same as Figure \ref{fig:halphadist}, but for the blue outliers.}
  \label{fig:wddist}
\end{figure}

We find a total of 389 blue outliers, 336 absorption line candidates and 1337 emission line candidates in our photometric catalogue. We show the magnitude and colour distributions of the emission and absorption line candidates in the left and right panels of Figure \ref{fig:halphadist}, respectively. We identify 62 sources that are both blue and have signs of excess H$\alpha$ emission. We also find 3 sources that are blue and candidate H$\alpha$ absorption sources. The sample properties of the blue outliers are shown in Figure \ref{fig:wddist}. 
The distribution of emission line candidates, in particular the decrease of the number of outliers in the bin 19.5\,$\leq r^{\prime}$\,$\leq$\,20, is introduced by the division into two magnitude bins at that $r^{\prime}$\,=\,19.5. The increase of H$\alpha$ emission line candidates up to $r^{\prime}$\,=\,19.5 can be explained by two effects: an increasing number of stars at fainter magnitudes, and an increased occurrence of emission line sources at fainter magnitudes. For the fainter magnitude bin, the intrinsic scatter of stars is much larger (due to the larger measurement errors), leading to a reduction in the number of observed outliers compared to the bright bin. 

Similarly, there is a large decrease in the number of absorption line candidates for $r^{\prime}$\,$\geq$\,19.5. This can be understood as a combination of two effects. As mentioned above, the increased scatter for faint stars decreases the number of outliers we identify. Moreover, absorption line candidates have a lower flux in the H$\alpha$ filter, hence  they are detected with a lower signal-to-noise ratio compared to other sources with similar $r^{\prime}$-band magnitudes. This means that it is much harder to detect absorption line candidates at fainter magnitudes, and results in a strong decrease in the number we can identify.

Table \ref{tab:outliers} shows an example of the information available for the outliers. The full tables can be found in the online material, and will also be made available in electronic form through the Vizier database (http://vizier.u-strasbg.fr). Here we will include the outlier catalogues described above, and in addition we will add outlier samples with slightly less restrictive selection critera. In particular, we will add an emission line catalogue which includes all 3$\sigma$ outliers and a catalogue of blue outliers with a 4$\sigma$ selection limit. 

\begin{table*}
 \centering
  \caption{Properties of the optical counterparts (identified as outliers) to GBS X-ray sources. The position of the optical counterpart is given in degrees. The uncertainties of these positions can be found in \citet{Wevers2016}. All magnitudes are given in the Vega system. The numbers in brackets correspond to the uncertainty on the last digit. EW is the equivalent width of the H$\alpha$ line in \AA, where a negative value indicates emission. In the comments we give the classification (if available) and the most relevant reference work. DN stands for dwarf nova, IP for intermediate polar, qLMXB for quiescent low mass X-ray binary, and Sy1 for Seyfert 1 galaxy.} 
 \begin{threeparttable}  
  \begin{tabular}{cccccccccccccc}
  \hline

CXID & RA ($^{\circ}$) & Dec ($^{\circ}$) & $r^{\prime}$ & $i^{\prime}$ & H$\alpha$ & $r^{\prime}$\,--\,$i^{\prime}$ & $r^{\prime}$\,--\,H$\alpha$ & EW&  $\frac{F_X}{F_{\text{opt}}}$ & Comments\\\hline
\multicolumn{3}{l}{Emission line sources} \\\hline
CX5&265.038086&--28.790512&19.03(2)&18.00(1)&18.03(2)&1.03&1.00&--50&  19.8& IP$^{a}$ \\
CX21&265.390747&--28.676245&19.14(2)&18.46(1)&17.37(2)&0.68&1.77&   &11.5&\\
CX37&264.371490&--29.467827&19.01(2)&18.35(2)&18.21(2)&0.66&0.80&--45&  6.4&IP$^{a}$ \\
CX81&266.109680&--27.323917&20.81(3)&19.95(3)&19.46(3)&0.86&1.35& &  16.7&DN$^{b}$ \\
CX93&266.186615&--26.058407&17.81(1)&16.66(1)&17.01(1)&1.15&0.80&--18.4&  0.73& CV$^{c}$ \\
CX118&264.709259&--28.802433&17.90(2)&16.94(1)&17.04(2)&0.96&0.86& &  0.81&\\
CX142&266.015655&--31.384815&21.23(3)&20.21(2)&20.28(5)&1.02&0.95& --59 &  13.5&DN$^{d}$ \\
CX207&266.606201&--26.526419&20.02(2)&19.25(3)&19.16(3)&0.77&0.86&--83&  4.4&IP$^{d}$ \\
CX585&265.953796&--31.416586&19.95(2)&19.23(1)&18.65(2)&0.72&1.30&  & 1.9 &\\
CX645&266.639374&--26.387234&19.43(2)&18.90(2)&18.77(2)&0.53&0.66& &  1.4& CV/qLMXB$^{b}$ \\
CX982&267.191925&-30.660831&21.9(1)&20.02(8)&19.96(5) & 1.9&2.0 & & 0.24 & DN?$^{b}$\\
CX1061&265.886413&-26.750890&18.77	(2)&16.77(1)&17.72(2)&2.00&1.05& & 0.01& \\
CXB279&266.340759&--32.160236&18.60(2)&16.80(1)&17.76(2)&1.80&0.84&  & 0.17&\\\hline
\multicolumn{3}{|l|}{Absorption line sources} \\\hline
CX2&264.368256&--29.133940&18.38(2)&16.78(1)&17.96(2)&1.60&0.42&--490(5) &  86& Sy1$^{e}$ \\
\multicolumn{3}{|l|}{Blue sources} \\\hline
CX361&267.781891&--29.677025&17.45(1)&17.64(2)&17.29(1)&-0.19&0.16&   & 0.63& AM CVn$^{f}$\\
CXB34&266.870453&--32.244946&21.39(4)&20.97(7)&21.34(1)&0.42&0.05& &  3.3&\\
  \hline
  \end{tabular}
  \begin{tablenotes}
  \small \item $^{a}$\citet{Britt2013}, $^{b}$\citet{Britt2014}, $^{c}$\citet{Ratti2013}, $^{d}$\citet{Torres2014}, $^{e}$\citet{Maccarone2012} $^{f}$\citet{Wevers2016b}
  \end{tablenotes}
  \end{threeparttable}
  \label{tab:outlierctparts}
\end{table*}

\subsection{Optical counterparts to X-ray sources}
\label{sec:xraymatches}
\begin{figure*}
  \includegraphics[height=12cm, keepaspectratio]{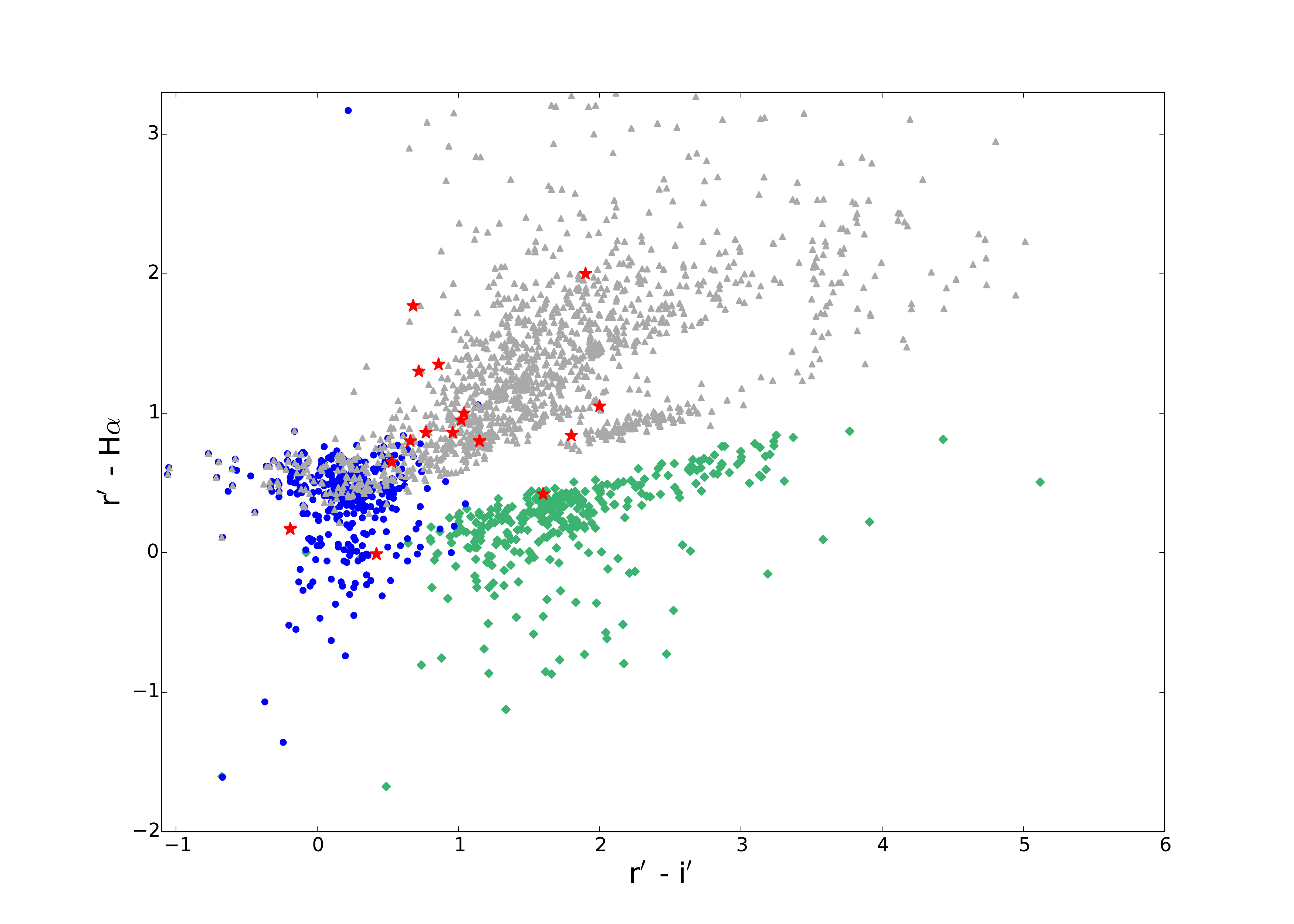}
  \caption{Colour-colour diagram of the three samples of outliers. Blue outliers are shown as blue circles, while emitter candidates are shown as grey triangles and absorption line candidates as green diamonds. Red stars indicate counterparts to X-ray sources in Table \ref{tab:outlierctparts}.}
  \label{fig:ccdoutliers}
\end{figure*}

Now that we have identified the outlier candidates in the optical photometry, we cross-correlate the sample with the GBS X-ray sources from \citet{Jonker2014}. We use the 4$\sigma$ X-ray error circle, defined in \citet{Wevers2016}, as the crossmatching radius. We find thirteen matches among the emission line candidates, while for the absorption line candidates we find one potential counterparts and among the blue outliers we find two matches. In Table \ref{tab:outlierctparts} we present the optical counterparts. 

\section{Discussion}
\label{sec:discussion}
\subsection{Optical counterparts to X-ray sources}
A significant fraction of the optical counterparts to X-ray sources found in this work have already been classified. In this regard, of the thirteen emission line candidates, seven have been previously studied and classified as CVs, three of which are magnetic systems (Table \ref{tab:outlierctparts}). This is not surprising, since magnetic systems are known to have a higher X-ray to optical flux ratio, so they can be detected at larger distances (in X-rays) compared to non-magnetic CVs. In addition, two H$\alpha$ emission line objects were identified as dwarf novae, while the remaining two are a dwarf nova candidate and either a CV or a qLMXB, respectively. One background AGN was identified as an H$\alpha$ absorption line candidate, and one blue outlier was classified as a high state AM CVn system.

\subsection{Multi-wavelength counterparts}
A number of observing programs with arcsecond spatial resolution have imaged the Galactic bulge, producing source catalogues which are potentially useful to further constrain the nature of objects in our sample. Below, we give a brief overview of the wavelength coverage and properties of some of these catalogues. A detailed study of multi-wavelength crossmatches with our catalogues is beyond the scope of this paper.

At optical wavelengths, the OGLE survey \citep{Udalski2015} has observed a large part of the Galactic bulge during multiple observing seasons with a cadence on occasion as short as 20 min, adding temporal information to our optical colours. In the near future VPHAS+ \citep{Drew2014} will complement our obervations by adding catalogues in the u$^{\prime}$, g$^{\prime}$ and z$^{\prime}$ bands, as well as temporal colour and astrometric information on a $\sim$\,10 year baseline. For some sources, proper motion measurements are available (e.g. \citeauthor{Sumi2004} \citeyear{Sumi2004}, \citeauthor{Fedorov2009} \citeyear{Fedorov2009}). Narrow-band He\,\textsc{i} 5875\,\AA\ information is also available from the UV EXcess survey \citep{Groot2009}. Other catalogues that overlap include the XMM-Newton serendipitous UV survey catalogue \citep{Page2012}, while the Vista Variables in the Via Lactea survey \citep{Saito2012} and UKIDDS \citep{Lucas2008} surveys can constrain the NIR part of the spectral energy distribution. Part of the GBS footprint coincides with mid-IR catalogues such as the Spitzer Bulge catalogue \citep{Uttenthaler2010}, the GLIMPSE catalogue \citep{GLIMPSE2009} and AllWISE catalogues \citep{Cutri2014}. 

\subsection{Colour-colour diagram of outliers}
We discuss the CCD of all the outliers and the implications for the identification of different source classes in the diagram. We show the position of all outliers in the CCD in Figure \ref{fig:ccdoutliers}. 
The emission line candidates are plotted as grey triangles, while blue outliers are marked as blue circles and absorption line candidates are shown as green diamonds. Counterparts to GBS X-ray sources are marked as red stars. 
The gap between emission and absorption line sources marks the global position of the locus of objects in all GBS fields. 

\begin{table}
 \centering
  \caption{Source identifications from the SIMBAD database. We used a search radius of 2 arcsec around the position of the optical source in question (H$\alpha$ emission or absorption line or blue outlier). The right ascension and declination of the GBS sources are given in degrees. Type gives the identification: X stands for X-ray source, RR signifies RR Lyrae star, DN means dwarf nova, YSO? and AGB? indicate a young stellar object candidate and asymptotic giant branch candidate, respectively. Em stands for emission line object, PN for planetary nebula, DQ for DQ Her type CV, and SRV for semi-regular variable star.}
  \begin{tabular}{cccc}
  \hline
RA & Dec. & Identifier & Type   \\\hline
\multicolumn{3}{l}{Emission line sources} \\\hline
266.780853&--25.958532 & IRAS 17440-2556 & Star  \\
267.191925&--30.660831&[JBN2011] 982& X \\
265.390747&--28.676244 & [SBM2001] 17 & X \\
264.796936&--28.320230 & BLG-RRLYR-24485 & RR \\
266.366790&--26.984680 &BLG-RRLYR-28521 & RR\\
264.428741&--29.110717 & BLG-RRLYR-23423 & RR \\
264.406219&--28.798717 & BLG-RRLYR-23355 & RR \\
267.461700&--30.052959&G359.5197-01.3648 & YSO? \\
266.377410&--31.789356 & J17453058-3147218 & YSO? \\
267.439086&--31.282707 & [KW2003] 64	 & Em  \\
267.468689&--30.550682&RPZM 42 & PN \\
266.109680&--27.323917 & [JBN2011] 81& DN  \\
266.186645&--26.058403&[JBN2011] 93& CV \\
265.067108&--29.060564&J174009.1-284725 & DQ \\
267.973144&--29.514860&J17515355-2930535 & Mira \\
268.338531&--28.346456 & J17532125-2820472 & AGB?\\
266.039398&--27.536483&J17440945-2732112& AGB? \\

\hline\multicolumn{3}{l}{Absorption line sources} \\\hline
268.905029&--28.736925&BLG-RRLYR-32252 & RR  \\
266.671905&--25.780313&BLG-RRLYR-29043	 & RR  \\
267.649688&--29.453142&BLG-LPV 64810 & SRV  \\
\hline\multicolumn{3}{l}{Blue sources} \\\hline
264.502564&--28.779535&BLG-RRLYR-23644 & RR \\
268.614349&--28.834324& BLG-RRLYR-31830 & RR  \\
268.926148&--29.134855&BLG-RRLYR-32289& RR \\
266.124786&--27.344559&PBOZ 10 & PN \\
  \hline
  \end{tabular}
  \label{tab:simbad}
\end{table}

\subsubsection{Emission line sources}
We interpret the spread of the emission line candidates (in $r^{\prime}$\,--\,H$\alpha$) as two different populations. The largest population consists of the outliers from the unreddened MS, and constitutes the upper population of emitters. In addition to these, another track of outliers at lower $r^{\prime}$\,--\,H$\alpha$ with a shallower slope can be identified. These systems are likely reddened stars with H$\alpha$ in emission.

We expect the sources that have $r^{\prime}$\,--\,$i^{\prime} \leq 1$ and H$\alpha$ in emission (the region of overlap between the blue outliers and emitters) to be good (non-magnetic) CV candidates. At somewhat redder colours we expect other H$\alpha$ excess sources such as intermediate polars (IPs) and (quiescent) LMXBs. For example, NS LMXBs and long period BH LMXBs are more X-ray bright, hence they can be detected further out in X-rays and their optical colours will be reddened by interstellar dust extinction. Intrinsically more red objects such as flare stars, and other sources such as chromospherically active (binary) stars and early-type emission line stars are also expected to be found as H$\alpha$ emitters. Planetary nebulae, whose spectrum is dominated by emission lines with a low continuum contribution, should show up as very large H$\alpha$ excess sources up to the $r^{\prime}$\,--\,H$\alpha$\,=\,3.3 limit. 

Crossmatching these sources with the SIMBAD database yields a variety of sources (Table \ref{tab:simbad}), indicating that indeed our sample could span the whole range of stellar evolution. We find, among others, two young stellar object candidates, five variable stars, two asymptotic giant branch candidates and four systems containing a WD.

\subsubsection{Blue sources}
\label{sec:bluesources}
The population of blue outliers likely consists of systems containing a compact object such as a white dwarf, neutron star or black hole. Nearby early-type MS stars are saturated in our observations, hence they cannot populate the blue part of the CCD. Although some outliers have bluer colours than field stars in the CCD, their colours are not extremely blue. The median ($r^{\prime}$\,--\,$i^{\prime}$) colour is 0.26, and because $r^{\prime}$\,--\,$i^{\prime}$ increases with distance (due to interstellar dust extinction) we expect the blue sources to be relatively close to Earth. The differential reddening in the $r^{\prime}$-band with respect to $i^{\prime}$ is $\Delta (r^{\prime}$\,--\,$i^{\prime})$ = 0.26\,$\times$\,A$_{r^{\prime}}$ \citep{Schlegel1998}, where A$_{r^{\prime}}$ is the extinction in the $r^{\prime}$-band (in mag). Assuming an intrinsic colour of $r^{\prime}$\,--\,$i^{\prime}$\,=\,0 (corresponding to an A0V spectral type, and typical for a blackbody spectrum with T\,$\sim$\,10000 K), this implies a typical reddening of A$_{r^{\prime}}$\,$\sim$\,1 mag which in turn means that the source should be nearby $\lesssim$ 1 kpc.

Blue outliers are identified up to $r^{\prime}$\,--\,$i^{\prime}$\,=\,1, indicating that we are not just finding foreground objects but also systems at distances where the dust extinction becomes appreciable (A$_{r^{\prime}}$\,$\sim$\,4 mag). For example, in some lines of sight with low extinction background AGN may be part of the blue outlier population. Some of the (dust) extinction may also be intrinsic to the system. Hot subluminous dwarfs (sdO/B stars) are also expected to be found as blue outliers with respect to field stars. Typical hot subdwarf stars fainter than $g$\,=\,17 mag have distances exceeding 4 kpc \citep{Geier2011}, implying that these blue objects could appear as some of the reddest of the blue outliers. The sample of 62 blue outliers that are also identified as emission line candidates likely consists of nearby CVs whose optical spectrum is dominated by the accretion disk, and in addition PN are known to populate this part of colour space (e.g. \citeauthor{Corradi2008} \citeyear{Corradi2008}). Blue sources showing indications of strong H$\alpha$ absorption (3 sources in our sample) are likely H-dominated (DA) WDs, and as explained below, there could be some AGNs among that sample. From Figure \ref{fig:ccdoutliers} it is clear that there are more blue sources that show signs of excess H$\alpha$ absorption, namely those sources with $r^{\prime}$\,--\,H$\alpha$\,$\leq$\,0, but they were not identified  as absorption candidates by our algorithm (see Section \ref{sec:methods}).

\subsubsection{Absorption line sources}
\label{sec:discussionabsorption}
Regarding the absorption line candidates, we expect them to include late-type stars and variable stars such as Mira giants and asymptotic giant branch (AGB) stars. Late-type stars can have strong ZrO absorption bands which coincide with the H$\alpha$ filter (e.g. \citeauthor{Castelaz1997} \citeyear{Castelaz1997}). This deficit of flux in the H$\alpha$ filter relative to the $r^{\prime}$-band shifts the sources below the main locus of objects in the CCD. \citet{Wright2008} investigated the nature of extremely red objects discovered in the IPHAS survey, and found that they contain a large sample of C-rich and S-type AGB stars. Similarly, we expect red sources ($r^{\prime}$\,--\,$i^{\prime}$\,$\gtrsim$\,2.5) with an excess H$\alpha$ absorption signature ($r^{\prime}$\,--\,H$\alpha$\,$\lesssim$\,1) to comprise C-rich and S-type AGB stars. The range of $r^{\prime}$\,--\,H$\alpha$ colour index originates in the relative strengths of different molecular absorption bands (in particular TiO, VO and ZrO) for varying surface chemistries. 

\begin{figure} 
  \includegraphics[height=4.6cm, keepaspectratio]{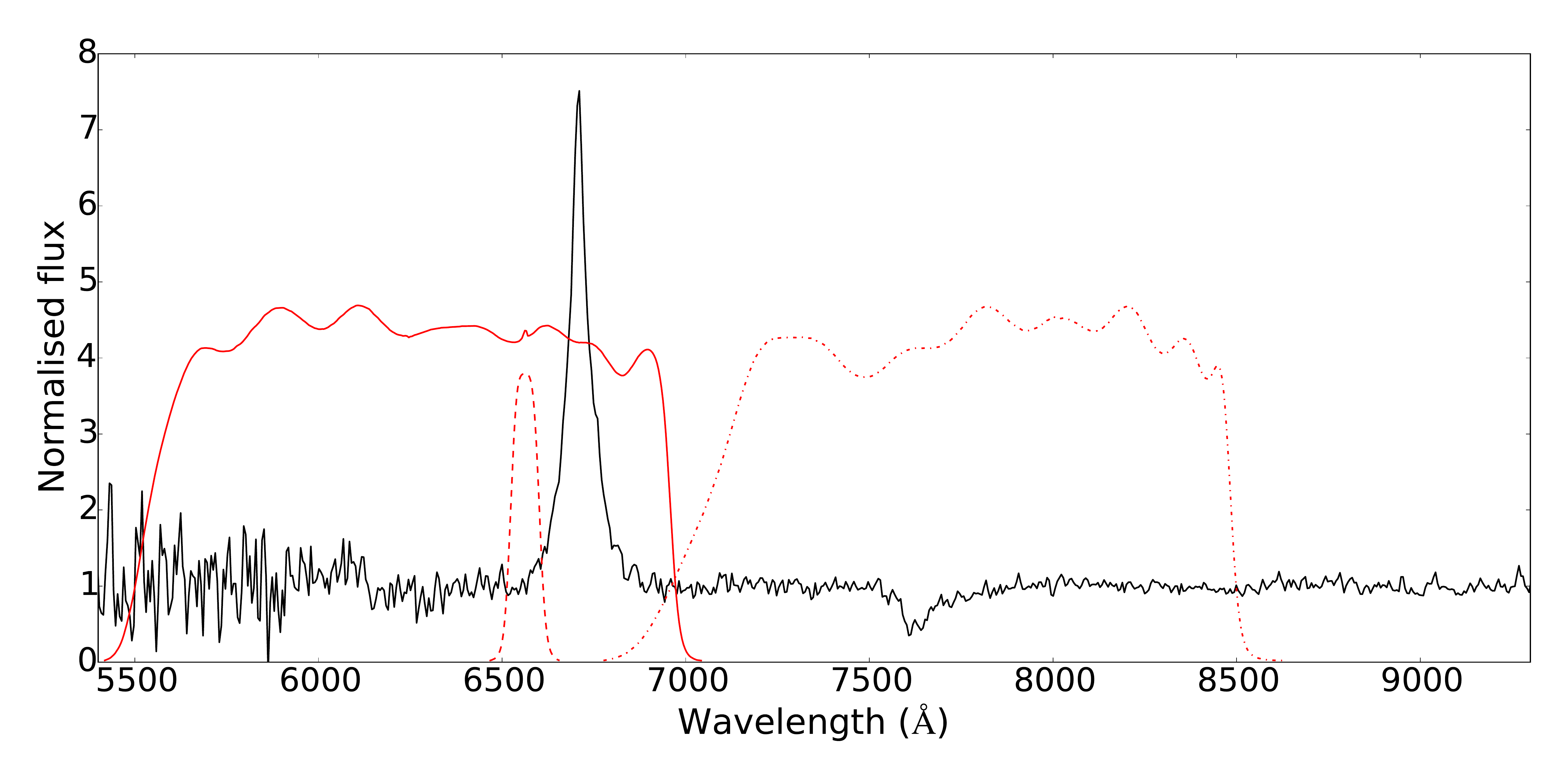}
  \caption{Continuum normalised EFOSC2 spectrum of CX2. The broad H$\alpha$ emission line is the only emission feature in the spectrum. Overplotted are the three filter profiles we have used in our photometric search: $r^{\prime}$ (left solid line), $i^{\prime}$ (dash-dotted line) and H$\alpha$ (dashed line).}
  \label{fig:cx2spec}
\end{figure} 

In addition to variable stars we expect another group of sources to populate this area in the CCD. We illustrate this using CX2, which was classified as a Seyfert 1 (Sy1) galaxy at a redshift of z\,=\,0.0214 \citep{Marti1998, Jonker2011, Maccarone2012}. In this work we have identified it as an H$\alpha$ absorption line source, which seems contradictory as Sy1 galaxies are known to have broad H$\alpha$ emission lines in their spectra. 
Figure \ref{fig:cx2spec} shows an EFOSC2 optical spectrum of CX2. Overplotted are the $r^{\prime}$, $i^{\prime}$ and H$\alpha$ filter profiles. The H$\alpha$ emission line of CX2 has been redshifted outside of the narrow H$\alpha$ filter bandpass and instead falls in the $r^{\prime}$ filter. Consequently, the source is brighter in $r^{\prime}$ relative to H$\alpha$, and falls in the region below the locus of sources in the CCD. 

An estimate of the redshift range for which this effect is at play can be obtained as follows. The width of our H$\alpha$ filter profile is $\sim$\,100\,\AA. Assuming an H$\alpha$ emission line FWHM in a Seyfert 1 galaxy of \,$\sim$\,100\,\AA$\ $\citep{Winkler1992}, this effect will cause AGNs with redshifts between z\,=\,0.015 and z\,=\,0.06 to potentially show up as absorption line sources (the exact boundaries depending on the FWHM of the emission line).
We thus expect that part of the background AGN population with a strong H$\alpha$ emission line could show up in our photometric search as candidate absorption line sources. More generally speaking, different subclasses of AGN exhibit multiple strong emission features (e.g. H$\beta$, [O\,\textsc{ii}] $\lambda3727$ and [O\,\textsc{iii}] $\lambda5007$) that, if at the right redshift, can fall into the $r^{\prime}$-band. This means that there are multiple redshift ranges, depending on which strong emission features are present in the spectrum, where the source could appear as an H$\alpha$ absorption line object. Similarly, AGNs with the right redshift may show up as H$\alpha$ emission line objects due to emission lines other than H$\alpha$.


\section{Summary}
\label{sec:summary}
We use optical photometry in three filters ($r^{\prime}$, $i^{\prime}$ and H$\alpha$) to create ($r^{\prime}$\,--\,$i^{\prime}$, $r^{\prime}$\,--\,H$\alpha$) colour-colour diagrams of point sources in the Galactic Bulge Survey fields \citep{Jonker2011, Jonker2014, Wevers2016}. The optical source catalogue reaches a mean 5$\sigma$ depth of $r^{\prime}$\,=\,22.5, $i^{\prime}$\,=\,21.1. The CCDs are used to systematically search for outliers in colour space, specifically H$\alpha$ emission and absorption line sources. We also use the $r^{\prime}$--$i^{\prime}$ colour distribution to search for blue outliers with respect to field stars. We identify 1337 emission line candidates, 336 absorption line candidates, and 389 blue outliers in the catalogue. These samples likely contain a plethora of sources, ranging from chromospherically active stars, subluminous hot dwarfs, white dwarfs, CVs, planetary nebulae, LMXBs and variable stars to background AGN. There is overlap between the blue and emission line candidates in 62 objects, and 3 blue sources are identified as having excess H$\alpha$ absorption. We crossmatch our outlier samples with the catalogue of GBS X-ray sources \citep{Jonker2014}, and find that sixteen outliers are counterparts to X-ray sources. Ten of those were previously classified based on photometric and/or spectroscopic follow-up. Four emission line candidates are classified as (magnetic) CVs, two as dwarf novae (and one DN candidate), and one system is a CV/qLMXB candidate. One of the absorption line candidates is a background AGN, and one blue outlier was classified as a high state AM CVn system. Spectroscopic observations of a representative sample of sources are needed to determine the completeness and sensitivity to the EW of our method. Individual source classifications require spectroscopic observations. The panchromatic coverage of the GBS area, including X-ray, UV, optical and IR observations, can greatly facilitate a targetted search for specific source classes.

\section*{Acknowledgements}
PGJ acknowledges support from European Research Council Consolidator Grant 647208. COH acknowledges support from an NSERC Discovery Grant, and Discovery Accelerator Supplement. We thank Tom Marsh for developing the software package \textsc{molly}. We thank N. Wright for useful suggestions about the nature of absorption line candidates. Based on observations collected at the European Organisation for Astronomical Research in the Southern Hemisphere under ESO programme 085.D-0441(C).


\bibliographystyle{mnras.bst}
\bibliography{bibliography_halpha.bib}

\label{lastpage}

\end{document}